\documentclass[%
 aip,
 amsmath,amssymb,
 reprint,%
]{revtex4-1}
\usepackage{color}%
\usepackage{graphicx}
\usepackage{dcolumn}
\usepackage{bm}

\usepackage[utf8]{inputenc}
\usepackage[T1]{fontenc}
\usepackage{mathptmx}
\usepackage{etoolbox}

\makeatletter
\def\@email#1#2{%
 \endgroup
 \patchcmd{\titleblock@produce}
  {\frontmatter@RRAPformat}
  {\frontmatter@RRAPformat{\produce@RRAP{*#1\href{mailto:#2}{#2}}}\frontmatter@RRAPformat}
  {}{}
}%
\makeatother
\begin{document}

\preprint{AIP/123-QED}

\title[Sample title]{A construction method of the quasi-monolithic compact interferometer based on UV-adhesives bonding}

\affiliation{MOE Key Laboratory of Fundamental Physical Quantities Measurements, The School of Physics, Huazhong University of Science and Technology, Wuhan 430074, China
}%
\author{Xiang Lin}
\author{Hao Yan$^*$}%
 \email{yanhao2022@hust.edu.cn}
\author{Yiqiu Ma}%
\author{Zebing Zhou}%

\date{\today}

\begin{abstract} 
Quasi-monolithic interferometers play a crucial role in high-precision measurement experiments, including gravitational wave detection, inertial sensing, vibrometry, and seismology. Achieving high stability and accuracy in such interferometers requires a method for bonding optical components to a baseplate. While optical contact bonding and silicate bonding are common methods, UV adhesives offer advantages such as controlled curing and low geometrical requirements for optical components and baseplates. This paper presents a detailed construction method for a quasi-monolithic compact interferometer based on UV-adhesive bonding. We built two types of interferometers using this method: a $100\,{\rm mm} \times 100\,{\rm mm}\times 20\,{\rm mm}$ Mach-Zender homodyne interferometer with unequal arm lengths of about $100\,{\rm mm}$ for laser frequency noise monitoring, and a heterodyne interferometer as a displacement sensing head sizing $20\,{\rm mm} \times 30\,{\rm mm}\times 20\,{\rm mm}$. Our Mach-Zender interferometer achieved a phase noise level of $2\,\mu{\rm rad}\sqrt{{\rm Hz}}$ at $1\,{\rm Hz}$ and a equivalent laser frequency noise monitoring sensitivity of about $1\,{\rm kHz}/\sqrt{{\rm Hz}}$ at $1\,{\rm Hz}$. The compact heterodyne interferometer sensing head showed a sensitivity level of $1\,{\rm pm}/\sqrt{{\rm Hz}}$ in translation and $0.2\,{\rm nrad}/\sqrt{{\rm Hz}}$ in two tilts above $0.4\,{\rm Hz}$. Our tests demonstrate that quasi-monolithic compact interferometers based on UV-adhesive bonding can achieve high sensitivity levels at the pico-meter and nano-radian scales.
\end{abstract}

\maketitle


\section{\label{sec:level1}INTRODUCTION}
High-precision compact interferometers find applications in various fields such as gravitational wave detectors,\cite{armano2021} inertial sensors,\cite{zhang2022quasi, isleif2019compact, hines2023compact, carter2020high} and vibrometers.\cite{yan2022all, wan2009monolithic, mahadevan2008inexpensive} For these applications, sensitivity requirements are often at the sub-nanometer or even picometer level. Opto-mechanical stability is crucial for such interferometers. Not only should the material of the baseplate and the optical components be ultra-stable, but the contact between them should also be ultra-stable.\cite{dehne2012construction}

The opto-mechanical stability of the interferometer bench mainly depends on the coefficient of thermal expansion and the mechanical stiffness. An optical bench with a low coefficient of thermal expansion and high mechanical stiffness is the key to achieving a high-precision interferometer.\cite{wan2011development} Conventional interferometers typically use commercial metal mirror mounts made of aluminum, stainless steel, or invar alloys, which have a relatively large coefficient of thermal expansion of about $10^{-5}-10^{-6}\,/{\rm K}$. Furthermore, the mirror mounts that use screws and springs to facilitate positional adjustment and reuse of optical components are further reduced in mechanical stiffness and thermal stability. In pursuit of higher accuracy and stability, interferometers tend to use quasi-monolithic designs, i.e., purely optical materials and contact methods.\cite{dehne2012construction}

One such technique is the optical contact bonding that joins two highly polished glass surfaces purely by van der Waals forces.\cite{greco2001optical} However, the main limitation is the low and unstable breaking strength, making it unsuitable for industrial metrology and space science. The second optical bonding technique is known as silicate bonding,\cite{van2008hydroxide, van2014hydroxide} which was successfully used on the optical bench of the LISA Pathfinder mission with fm-level accuracy.\cite{armano2021, schwarze2019picometer} The basic principle is to use chemical reactions to break and reconnect the Si-O bond between the optical components and the bench, resulting in excellent mechanical stiffness and thermal stability. However, the main disadvantage is the short chemical reaction time (about tens of seconds) left for position and attitude adjustments of the optical components, resulting in complex manufacturing processes and high cost. In addition, this method requires two highly polished glass surfaces to be bonded.

Another optical bonding method is adhesive bonding, i.e., UV adhesives or epoxy resin adhesives, etc.\cite{yi2017laboratory, preston2012quasi} The robustness of the two-component epoxy resin adhesive bonds against mechanical and thermal inputs has been tested, \cite{ressel2010ultrastable} making this new bonding technique a potential option for interferometric applications in future space missions. Epoxy adhesives have a relatively high bond strength, but still require a long curing time, waiting approximately two days for a single adhesive to cure. The bonding process with UV adhesives can be controlled by UV light and takes only a few tens of seconds with a UV lamp. Compared to optical contact and silicate bonding, UV-adhesive bonding has a longer time scale for alignment, as well as shorter curing times, which is convenient for the manufacture of quasi-monolithic interferometers. The main disadvantages of UV adhesive are the relatively poor bond strength and thermal stability.

The quasi-monolithic interferometer construction method presented in the paper uses UV adhesive bonding and low thermal expansion materials like fused silica to achieve stable and robust bonds between the optical components and the baseplate. The construction process includes the construction of a quasi-monolithic fiber collimator, positioning and pointing adjustment scheme of optical components, and the UV-adhesive bonding process.

Unequal arm length control is critical for interferometers to achieve common-mode rejection of ambient noise or laser frequency noise monitoring. A wide-range swept laser is used in the setup to monitor unequal arm length with sub-millimetre accuracy during the position tuning and assembly process of the optical components.

The Mach-Zender homodyne interferometer built using this construction method has a large unequal-arm length and can be used for noise monitoring, including laser frequency noise and temperature fluctuations. The heterodyne interferometer built with this method can achieve picometre or nano-radian levels of displacement measurement. 

\section{\label{sec:level2}EXPERIMENTAL KEY TECHNOLOGIES}
There are three key technologies for constructing a quasi-monolithic glass interferometer based on UV-adhensive bonding: (1) the construction of the quasi-monolithic fiber collimator, (2) the positioning and pointing adjustment of the optical components (fiber collimator, beam splitter, reflecting mirror) on the baseplate, and (3) the UV bonding process. The details are presented as follows.

\subsection{\label{sec:level 2.1} Quasi-monolithic glass fiber collimator}
A fiber optic collimator plays a crucial role in an interferometer by converting light from a fiber into collimated spatial light. It typically consists of a fiber ceramic ferrule, one or more lenses, and an external holder. However, commercial collimators often use metal mounts to adjust the focus and attitudes of the lenses, making them unsuitable for constructing all-glass interferometers. Alternatively, some glass collimators bond the fiber ceramic ferrule and the lens inside a glass ferrule using a C-lens or G-lens.

\begin{figure}
\includegraphics[width=8cm]{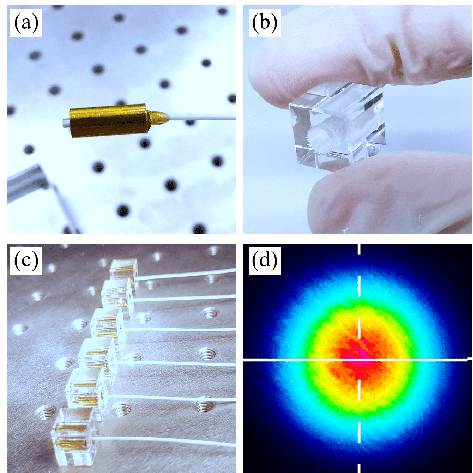}
\caption{\label{fig:1} The quasi-monolithic glass fiber collimator. (a) is the G-lens with a size of $4\,{\rm mm}$ in diameter and $8\,{\rm mm}$ in length, (b) is the $10\,{\rm mm}$ cubic glass mount with a cylindrical hole in the center, (c) shows the quasi-monolithic glass fiber collimators, and (d) shows the optical power distribution of the laser emitted from the quasi-monolithic glass fiber collimator.}
\end{figure}

To create a fiber collimator suitable for quasi-monolithic glass interferometers, we used a small custom-made glass cube with a cylindrical hole as a bonding mount for the G-lens. The G-lens and the cubic glass mount were bonded using UV adhesives to form our quasi-monolithic glass fiber collimator, as shown in Fig.~\ref{fig:1}. The outgoing laser beam from this collimator was found to have a beam waist of approximately $0.2\,{\rm mm}$, corresponding to a Rayleigh distance of about $0.12\,{\rm m}$. The optical power distribution of the laser emitted from the quasi-monolithic collimator followed a Gaussian distribution with an ellipticity of $0.99$, as shown in Fig.~\ref{fig:1}(d).

\subsection{Optical alignment}\label{sec:level 2.2}
In constructing a quasi-monolithic glass interferometer, precise positioning and angular alignment of optical components are essential. There are several reasons for this, including achieving high contrast interference, reducing cosine errors, and controlling unequal arm lengths. The latter is especially important because smaller unequal arm lengths lead to lower interferometer noise.

\begin{figure}
\includegraphics[width=8cm]{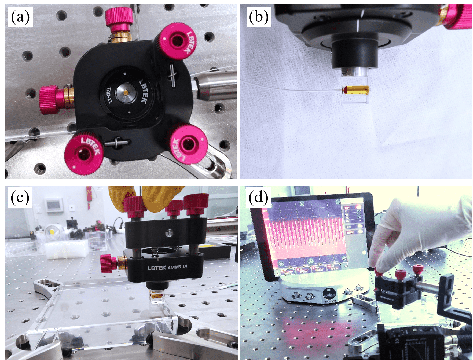}
\caption{\label{fig:2} The positioning and pointing adjustment scheme. (a) shows the six degree-of-freedom mirror mount, (b) shows the quasi-monolithic glass collimator on the mirror mount, (c) shows the position and attitude adjustment of the optical components, (d) shows the allignment of two laser beams by maximising their overlap to achieve good interferometric contrast.}
\end{figure}

To facilitate precise alignment, a six-degree-of-freedom mirror mount is used in the setup, as shown in Fig.~\ref{fig:2}. This mirror mount has three degrees of freedom of translation and three degrees of freedom of rotation, allowing for sub-millimeter and sub-milliradian adjustments. The main steps of optical component alignment are as follows: (1) fix the target component (collimator, beamsplitter, reflector, etc.) on the six-degree-of-freedom mirror mount as in Fig.~\ref{fig:2}(b), (2) place the target component in an approximately suitable position in the optical system according to the optical path design, (3) adjust the height of the target component from the substrate to a few millimeters, (4) adjust the remaining five degrees of freedom of the target component until the positions and attitudes are as designed as in Fig.~\ref{fig:2}(c), (5) lower the height of the target component to the substrate.

Among the various optical components, the positioning and pointing of the final component (typically a beam splitter) responsible for the interferometric beam combination are crucial. It directly impacts the interference contrast and requires high-precision adjustment. In contrast, the positioning and pointing accuracy requirements of other optical components are relatively lower. During the optical bench construction, non-critical optical elements are adjusted with the aid of a small aperture diaphragm to ensure that all laser beams are in the same plane. The final beam combination is adjusted by observing the interference contrast as feedback, as shown in Fig.~\ref{fig:2}(d).

\subsection{\label{sec:level 2.3} UV-adhesives bonding}
The UV adhesive bonding process is a critical step in the construction of quasi-monolithic glass interferometers. In comparison to other bonding methods such as optical contacting bonding and silicate bonding, UV adhesive bonding is more convenient and suitable for situations where high mechanical rigidity is not required.

\begin{figure}
\includegraphics[width=8cm]{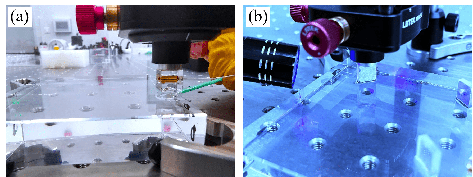}
\caption{\label{fig:3} UV-adhesives bonding process. (a) shows the UV-adhesives dispensing with a special stick in the gap between the baseplate and the optical component, (b) shows the process of curing the adhesives with a UV lamp.}
\end{figure}

Fig.~\ref{fig:3} illustrates the UV adhesive bonding process. Initially, the target component is positioned and adjusted with a large gap to the baseplate using the 5-Dofs adjustment mount. Then, UV adhesive is dispensed using a special stick and applied evenly while ensuring the correct amount is used. The target component is then lowered to fit the upper surface of the optical baseplate while maintaining its position and attitudes. Finally, the optics are illuminated with a UV lamp until the adhesive has cured. This process is repeated until all the optical components are adhesive-bonded in their designed positions and attitudes.

\section{\label{sec:level 3}INTERFEROMETER CONSTRUCTION}
Using the techniques described above, we have successfully constructed two classic quasi-monolithic glass interferometers. The first one is a homodyne Mach-Zender interferometer with a large unequal arm length, which is suitable for noise monitoring, including laser frequency noise and temperature fluctuations. The second one is a compact heterodyne interferometer with an optical path configuration consisting of two symmetrical Michelson structures. When combined with differential wavefront sensing (DWS), it enables high-precision measurements in three degrees of freedom, including one translation and two tilts.

\subsection{\label{sec:level 3.1} Mach-Zender interferometer}
\begin{figure}
\includegraphics[width=6cm]{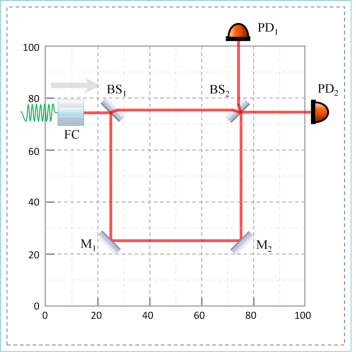}
\caption{\label{fig:4} Schematic design of the optical path of the homodyne Mach-Zender interferometer with a unequal arm length of about $100\,{\rm mm}$. FC: fiber collimator, BS: beam splitter, PD: photo detector.}
\end{figure}

\begin{figure}
\includegraphics[width=8cm]{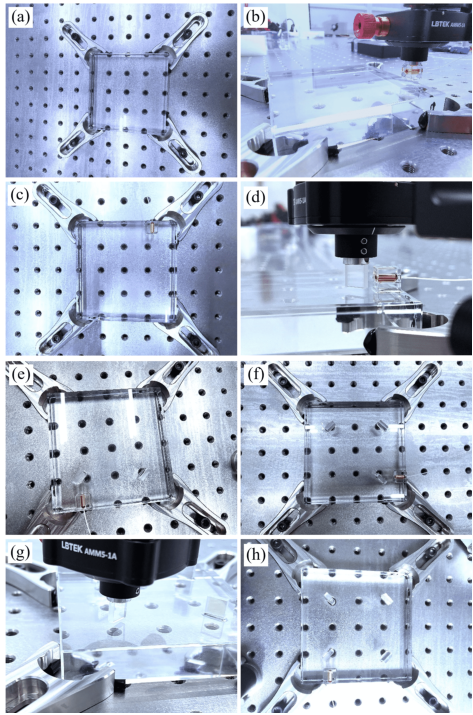}
\caption{\label{fig:5} UV-adhesives bonding process of the quasi-monolithic glass homodyne Mach-Zender interferometer. (a) is the fused-silica bench with a size of $100\,{\rm mm} \times 100\,{\rm mm}\times 20\,{\rm mm}$, (b) and (c) show the adjustment of the quasi-monolithic glass collimator, (d)-(g) show the UV bonding process of the remaining optical components (two beam splitters and two reflectors), (h) shows the finished quasi-monolithic glass Mach-Zender interferometer.}
\end{figure}

\begin{figure}
\includegraphics[width=6cm]{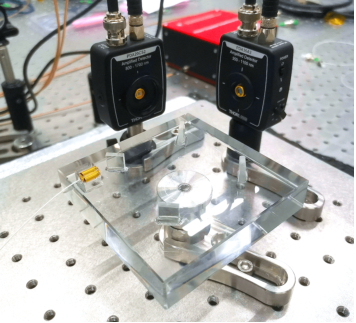}
\caption{\label{fig:6} The test of the completed quasi-monolithic glass Mach-Zender interferometer.}
\end{figure}

Conventional Mach-Zender interferometer is of equal arm length. 
In order to achieve unequal-arm interference, a slight modification was made to the optical path design of a conventional Mach-Zender interferometer, as shown in Fig.~\ref{fig:4}. The laser beam is injected from the left and divided into two paths. The upper short path transmits directly, while the lower long path is reflected twice by two mirrors (${\rm M_1}$ and ${\rm M_2}$) before interfering with the upper path. Finally, the interference signal is balanced by a pair of photodetectors (${\rm PD_1}$ and ${\rm PD_2}$). The unequal arm length is approximately $100\,{\rm mm}$ according to the drawing design.

Figure~\ref{fig:5} shows the UV-adhesive bonding process of the quasi-monolithic glass homodyne Mach-Zender interferometer. The optical baseplate and components are made of fused silica with a thermal expansion coefficient of about $0.5\times10^{-6}\,{\rm /K}$. The completed quasi-monolithic glass Mach-Zender interferometer, with dimensions of $100\,{\rm mm}\times100\,{\rm mm} \times 30\,{\rm mm}$, was tested and shown in Fig.~\ref{fig:6}. The adjustment method shown in Fig.~\ref{fig:2} is effective. After careful adjustment of the position and attitude of the optical components, the laser frequency sweep test showed an interference contrast of over $0.8$, indicating that the laser beams were well coincided and aligned.

The laser frequency noise can be measured by an unequal-arm interferometer.~\cite{gerberding2017laser} The measurement equation is given by:
\begin{equation}
\delta f=\frac{c}{2\pi \Delta L}\delta \varphi, 
\label{eq:1}.
\end{equation}
where $\delta f$ is the laser frequency noise, $c$ is the speed of light, and $\Delta L$ is the unequal armlength of the Mach-Zender interferometer. The frequency-phase coefficient $c/{2\pi \Delta L}$ is about $500\,{\rm Hz}/\mu {\rm rad}$ with a unequal armlength of about $100\,{\rm mm}$ in our setup.

\subsection{\label{sec:level 3.2} Heterodyne interferometer}

In addition to the homodyne Mach-Zender interferometer, we also developed a compact heterodyne interferometer that is ideal for large dynamic measurements due to its linear response to the translational displacement of the target reflector. Moreover, the differential wavefront sensing techniques utilized in this interferometer allow for two additional tilt measurements.

\begin{figure}
\includegraphics[width=8cm]{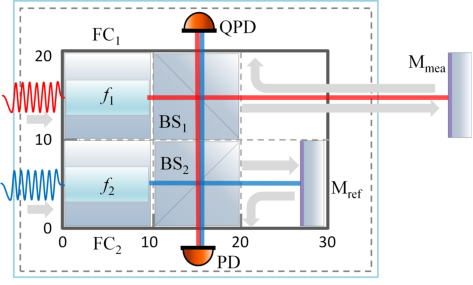}
\caption{\label{fig:7} Schematic design of the optical path of the compact heterodyne interferometer. QPD: quadrant photo detector, ${\rm M_{ref}}$: reference mirror, ${\rm M_{mea}}$: measurement mirror.}
\end{figure}

\begin{figure}
\includegraphics[width=8cm]{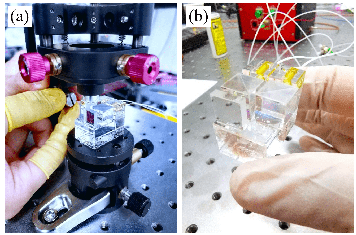}
\caption{\label{fig:8} Pictures of the compact heterodyne interferometer with a size of $20\,{\rm mm}\times 30\,{\rm mm}\times 20\,{\rm mm}$. (a) shows the adjustment of the optical components, (b) shows the finished compact bench.}
\end{figure}

To enhance the sensitivity of the interferometer, we designed two symmetrical Michelson-type optical path structures that suppress ambient noise coupling.\cite{yan2020highly, yan2015dual, S2014Compact, Xiangzhi2015Beam} The interferometer, as shown in Fig.~\ref{fig:7}, uses two laser beams ($f_1$ and $f_2$) with a fixed frequency difference, which are injected by collimators (${\rm FC_1}$ and ${\rm FC_2}$) from the left and split into two beams by two beam splitters (${\rm BS_1}$ and ${\rm BS_1}$), respectively. The transmitted beams hit the reference mirror (${\rm M_{ref}}$) and the target measurement mirror (${\rm M_{mea}}$), respectively, and are reflected back into the photodetector (PD and QPD) by ${\rm BS_1}$ and ${\rm BS_2}$. The reference beams directly enter the photodetector (PD and QPD). Finally, two interference beat signals are obtained: the upper one is the measurement signal (measured by QPD), and the lower one is the reference signal (measured by PD). The three degrees of freedom motion of the target mirror can be obtained by equations:
\begin{align}
&x=\frac{\lambda }{4\pi } \left (\frac{\varphi _{1}+\varphi _{2}+\varphi _{3}+\varphi _{4}}{4}-\varphi _{5} \right ),\\
&\theta _{\rm pitch} =k \left (\varphi _{1}+\varphi _{2}-\varphi _{3}-\varphi _{4}\right ),\\
&\theta _{\rm yaw} =k \left (\varphi _{1}+\varphi _{4}-\varphi _{2}-\varphi _{3}\right ),
\label{eq:4}
\end{align}
where $\{x,\theta _{\rm pitch},\theta _{\rm yaw}\} $ denotes the displacement of translation and two tilts (pitch and yaw), respectively, and $\lambda$ represents the wavelength of the laser. The four quadrant phase signal of the QPD is represented by $\{\varphi _{1},\ \varphi _{2},\ \varphi _{3},\ \varphi _{4}\} $, while $\varphi _{5}$ denotes the phase signal of the PD. The coefficient of the phase to the angle, denoted by $k$, requires experimental calibration and depends on the waist diameter, wavefront, PD size, etc. The compact heterodyne interferometer has a size of $20\,{\rm mm} \times 30\,{\rm mm} \times 20\,{\rm mm}$, as shown in Fig.~\ref{fig:8}.

\section{\label{sec:level 4} EXPERIMENTAL RESULTS}
\subsection{\label{sec:level 4.1} Mach-Zender interferometer}
\begin{figure}
\includegraphics[width=8cm]{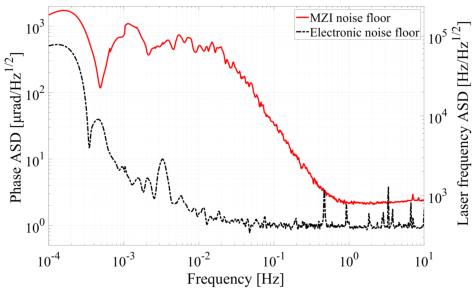}
\caption{\label{fig:9} Phase noise of the qutasi-monolithic glass MZI (Mach-Zender interferometer). The solid red curve shows the phase noise of the MZI inside a insulated tank. The dashed black curve shows the electronic noise floor, obtained by replacing the interferometric signal with an single laser beam for photoelectric detection. The coordinates on the right shows the corresponding equivalent laser frequency noise measurement sensitivity.}
\end{figure}

To test the phase noise floor of the quasi-monolithic glass Mach-Zender interferometer, we placed the optical bench inside a temperature-insulated tank to isolate temperature fluctuations and allow temperature modulation tests. We used a commercial laser with low power and frequency noise to inject light into the Mach-Zender interferometer. The sensitivity curve of the Mach-Zender interferometer is shown in Fig.~\ref{fig:9}, which achieved a phase noise floor of $2\,\mu{\rm rad}/\sqrt{{\rm Hz}}$ above $1\,{\rm Hz}$. According to Eqs.~(\ref{eq:1}), the laser frequency noise floor of $1\,{\rm kHz}/\sqrt{{\rm Hz}}$ above $1\,{\rm Hz}$ can be obtained with an unequal armlength of $100\,{\rm mm}$, as shown in the right coordinate.

\subsection{\label{sec:level 4.2} Heterodyne interferometer}
The quasi-monolithic glass heterodyne interferometer has the capability to measure motion in three degrees of freedom using differential wavefront sensing (DWS). Unlike the homodyne Mach-Zender interferometer, the heterodyne interferometer generates an AC beat signal which requires a phasemeter to measure. To evaluate the noise floor, a target reflector is attached to an optical stage and the heterodyne interferometer is used to measure its motion in three degrees of freedom. The translation and two tilts (pitch and yaw) can be obtained from the signals of the quadrant photodetector (QPD) and photodetector (PD) using Eq.~(\ref{eq:4}).

\begin{figure}
\includegraphics[width=8cm]{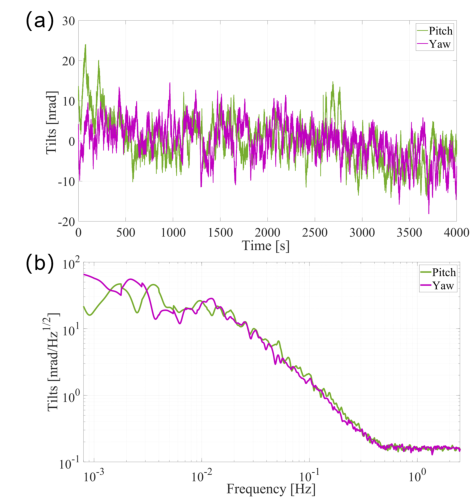}
\caption{\label{fig:10} Tilts noise of the quasi-monolithic compact heterodyne interferometer. (a) shows the time-domain noise curve without displacemnet input, and (b) shows the noise amplitude spectral density curve.}
\end{figure}

\begin{figure}
\includegraphics[width=8cm]{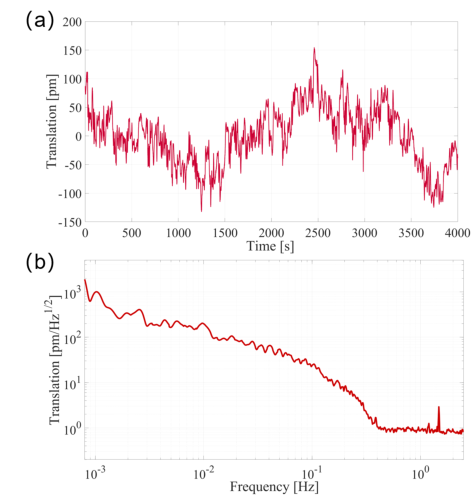}
\caption{\label{fig:11} Translation noise of the quasi-monolithic compact heterodyne interferometer. (a) is the time-domain curve, and (b) is the noise amplitude spectral density curve.}
\end{figure}

The noise floor for the two tilts is depicted in Fig.~\ref{fig:10}. A tilt sensitivity of $0.2\,{\rm nrad}/\sqrt{{\rm Hz}}$ is achieved above $0.4\,{\rm Hz}$, with the noise curve increasing rapidly and reaching $30\,{\rm nrad}/\sqrt{{\rm Hz}}$ at $0.01\,{\rm Hz}$. The situation is similar for translation measurement, as shown in Fig.~\ref{fig:11}. A translation sensitivity of $1\,{\rm pm}/\sqrt{{\rm Hz}}$ is achieved above $0.4\,{\rm Hz}$, with the noise curve increasing rapidly and reaching $200\,{\rm pm}/\sqrt{{\rm Hz}}$ at $0.01\,{\rm Hz}$.

\section{\label{sec:level 5} CONCLUSION}
We have successfully demonstrated a method for constructing quasi-monolithic compact interferometers using UV-adhesives, resulting in the creation of two optical benches. One of these benches is a Mach-Zender homodyne interferometer with unequal arm lengths of $100\,{\rm mm}$, which is capable of noise monitoring. The phase noise of this interferometer reaches $2\,\mu {\rm rad}/\sqrt{{\rm Hz}}$ at $1\,{\rm Hz}$, and its equivalent frequency monitoring sensitivity is about $1\,{\rm kHz}/\sqrt{{\rm Hz}}$ at $1\,{\rm Hz}$. The other bench is a 3-Dofs compact heterodyne interferometer with a sensitivity of $1\,{\rm pm}/\sqrt{{\rm Hz}}$ in translation and $0.2\,{\rm nrad}/\sqrt{{\rm Hz}}$ in two tilts (pitch and yaw) above $0.4\,{\rm Hz}$.

In comparison to other bonding methods such as optical contact bonding, silicate bonding, and epoxy resin adhesives bonding, UV-adhesives bonding offers advantages such as controlled curing time and low geometrical machining requirements for the optical components and baseplates, making it a suitable option for constructing quasi-monolithic interferometers at the pico-meter and nano-radian levels.

\begin{acknowledgments}
This work was supported by the National Natural Science Foundation of China (Grant Nos. 12105375).
\end{acknowledgments}

\section*{AUTHOR DECLARATIONS}
\subsection*{Conflict of Interest}
The authors have no conflicts to disclose.
\section*{Author Contributions}
\textbf{Xiang Lin}: Experiment; Investigation (equal);
\textbf{Hao Yan}: Conceptualization (equal); Investigation (equal);
Writing – original draft.
\textbf{Yiqiu Ma}: Conceptualization (equal); Writing – review \& editing (equal).
\textbf{Zebing Zhou}: Investigation (equal); Writing – review. 

\section*{Data Availability Statement}
The data that support the findings of this study are available from the corresponding author upon reasonable request.

\nocite{*}

\section*{REFERENCE}
\bibliography{ref}

\end{document}